\documentclass[12pt]{article}
\usepackage{verbatim,epsfig,amsmath,url,amssymb,amsfonts}

\usepackage[round]{natbib}

\usepackage[margin=1in]{geometry}
\date{}

\begin{document}
\title{Theoretical Properties and Practical Performance of Fully Robust One-Sided Cross-Validation}
\author{Olga Y. Savchuk, Jeffrey D. Hart}

\maketitle

\begin{abstract}
Fully robust OSCV is a modification of the OSCV method that produces consistent bandwidth in the cases of smooth and nonsmooth regression functions. The current implementation of the method uses the kernel $H_I$ that is almost indistinguishable from the Gaussian kernel $\phi$ on the interval $[-4,4]$, but has negative tails. The theoretical properties and practical performances of the $H_I$- and $\phi$-based OSCV versions are compared. The kernel $H_I$ tends to produce too low bandwidths in the smooth case. The $H_I$-based OSCV curves are shown to have wiggles appearing in the neighborhood of zero. The kernel $H_I$ uncovers sensitivity of the OSCV method to a tiny modification of the kernel used for the cross-validation purposes. The recently found robust bimodal kernels tend to produce OSCV curves with multiple local minima. The problem of finding a robust unimodal nonnegative kernel remains open.

\bigskip


\vspace{0.5cm}

\end{abstract}

\section{Introduction}

Nonparametric regression estimation involves selecting a smoothing parameter, usually called the {\it bandwidth}, that mainly determines the appearance of a regression estimate. Inappropriately small bandwidth results in a bumpy estimate that tracks almost every data point on the scatter diagram, whereas too large bandwidth produces an oversmoothed regression estimate that may fail to represent important features of the regression function such as multiple peaks, sharpness of a peak, etc. There exist many methods that use the data to estimate the bandwidth that is optimal in certain sense. The most frequently used data-based bandwidth selection methods are the plug-in rule of~\citet{RSW:plug-in} and the cross-validation (CV) method of~\citet{Stone:CV}. There are many variations of both methods. One of the successful modifications of the CV method is the one-sided cross-validation (OSCV) method developed by~\citet{HartYi}.

All original OSCV research relies on the assumption that the regression function is {\it smooth}, which means that it has at least two continuous derivatives. \citet{HartYi} showed that using OSCV instead of CV may produce up to twentyfold reductions of the asymptotic bandwidth variance. \citet{Yi:OSCVsims} conducted a simulation study to illustrate the improved stability of OSCV compared to CV in finite samples. \citet{HartLee:OSCVrobust} argued that OSCV is robust to moderate levels of autocorrelation. \citet{OSCV:density} developed a version of the OSCV method for the kernel density estimator.

For many real data sets in economics, medicine, biology and other fields, the relationship between the variables is described by a continuous function that has sharp corners or {\it cusps} appearing at the points where the first derivative of a function has simple discontinuities. A continuous regression function with cusps is refereed to as {\it nonsmooth}. The original OSCV method produces a biased estimator of the optimal bandwidth in the nonsmooth case.  \citet{Savchuk:OSCVnonsmooth} developed the {\em fully robust OSCV} version that results in a consistent estimation of the optimal bandwidth regardless of the regression function's smoothness.

This article provides a detailed investigation of the theoretical properties and practical performance of the current implementation of the fully robust OSCV method. We also demonstrates performance of OSCV based on our recently found robust bimodal kernel.

The rest of the article proceeds as follows. In Section 2 we overview the problem of nonparametric regression estimation and outline the steps in the OSCV method. Section 3 contains an extended discussion of the fully robust OSCV method and brings new light onto performance of the original OSCV version of~\citet{HartYi}. Section 4 contains summary of our findings. The appendix includes certain supplementary materials.

Our subsequent presentation requires introducing the following notation. For an arbitrary function $g$, define
\begin{equation}
\label{eq:R_mu_notation}
R_g=\int_{-\infty}^\infty g^2(u)\,du,\qquad\qquad \mu_{2g}=\int_{-\infty}^\infty u^2g(u)\,du,
\end{equation}
\begin{equation}
\label{eq:J}
J_g=\left(\frac{R_g}{\mu_{2g}^2}\right)^{1/5},
\end{equation}
\begin{equation}
\label{eq:B}
B_g=\int_0^1\left\{z\bigl(1-D_g(z)\bigr)+G_g(z)\right\}^2\,dz+\int_0^1\left\{zD_g(-z)+G_g(-z)\right\}^2\,dz,
\end{equation}
and for all $z$,
\begin{equation*}
\begin{array}{l}
\displaystyle{D_g(z)=\int_{-\infty}^z g(u)\,du},\\[0.3cm]
\displaystyle{G_g(z)=\int_{-\infty}^z ug(u)\,du}.
\end{array}
\end{equation*}

\section{Nonparametric regression estimation and the OSCV method}

In the nonparametric regression model the observations $Y_1,Y_2,\ldots,Y_n$ are assumed to be generated as
\[
Y_i=r(x_i)+\varepsilon_i,
\]
where $r$ is an unknown regression function defined on the interval $[0,a]$, $a>0$, and
$\varepsilon_1,\varepsilon_2,\ldots,\varepsilon_n$ are uncorrelated
error terms such that $E(\varepsilon_i)=0$ and
$\mbox{Var}(\varepsilon_i)=\sigma^2$, $i=1,\ldots, n$. The design points
$x_1<x_2<\cdots<x_n$ are assumed to be fixed quantiles of the design density $f$. In the case of an evenly spaced design, $f\equiv1/a$.

The OSCV method is intended to select the bandwidths of the Gasser-M\"{u}ller estimator (see~\citet{GasserMuller}) or the local linear estimator (see~\citet{Cleveland:LLE}). In this article we concentrate on the local linear estimator, defined at the point $x$ as
\begin{equation}
\label{eq:LLE}
\hat r_h(x)=\frac{\sum_{i=1}^n
w_i(x)Y_i}{\sum_{i=1}^n w_i(x)},
\end{equation}
where $h>0$ is the bandwidth,
\begin{equation}
\label{eq:w}
w_i(x)=K\left(\frac{x-x_i}{h}\right)\left(t_{n,2}-(x-x_i)t_{n,1}\right),
\end{equation}
and
\begin{equation}
\label{eq:t}
t_{n,j}=\sum_{i=1}^n K\left(\frac{x-x_i}{h}\right)(x-x_i)^j,\quad
j=1,2.
\end{equation}
The kernel function $K$ is of the second order, that is it integrates to one, has zero first moment, and finite second moment (see~\citet{Wand:KS}).

Some popular measures of closeness of $\hat r_h$ to $r$ are the mean average squared error (MASE) and the average squared error (ASE). The ASE function is defined as
\begin{equation*}
\mbox{ASE}_K(h)=\frac{1}{n}\sum_{i=1}^n\left(\hat
r_h(x_i)-r(x_i)\right)^2.
\end{equation*}
The subscript ``$K$'' is used above to emphasize dependance of $\hat r_h$ on the kernel $K$.
The bandwidth $\hat h_0$ that minimizes ASE is optimal for the data set at hand. The MASE function is defined as the expectation of the ASE function.
The bandwidth $h_0$ that minimizes MASE is optimal in the average sense for all data sets generated from $r$ at the fixed values of $n$ and $\sigma$.

In the case of a smooth regression function, the asymptotic MASE expansion for the local linear estimator $\hat r_h$ based on the kernel $K$ has the following form:
\begin{equation*}
\mbox{MASE}_K(h)=\mbox{AMASE}_K(h)+o\left(h^4+\frac{1}{nh}\right),
\end{equation*}
where
\begin{equation}
\label{eq:AMASE_smooth}
\mbox{AMASE}_K(h)=\frac{R_K\sigma^2}{nh}+\frac{\mu_{2K}^2h^4}{4}\int_0^a(r^{\prime\prime}(x))^2f(x)\,dx.
\end{equation}
The minimizer of the $\mbox{AMASE}_K(h)$ function is
\[
h_n=\left(\frac{R_K\sigma^2}{\mu_{2K}^2 \int_0^a
\left(r\,^{\prime\prime}(x)\right)^2f(x)\,dx}\right)^{1/5}n^{-1/5}=J_KC_{r,\sigma}n^{-1/5},
\]
where
\begin{equation}
\label{eq:C_rsigma}
C_{r,\sigma}=\left(\frac{\sigma^2}{\int_0^a
\left(r\,^{\prime\prime}(x)\right)^2f(x)\,dx}\right)^{1/5}.
\end{equation}

The OSCV method is designed to produce an estimate of the MASE-optimal bandwidth $h_0$.
The main idea behind OSCV is to use different kernels in the estimation and cross-validation stages. The final regression estimate $\hat r_h$ is computed by using the local linear estimator based on a highly efficient kernel $K$, such as Gaussian, Epanechnikov, etc. Kernel efficiency discussion may be found in~\citet{Wand:KS}. In the cross-validation stage, one uses a so-called one-sided estimator $\tilde r_b$, where $\tilde r_b(x_i)$ is a local linear estimator computed from the data points $(x_1,Y_1),\ldots,(x_i,Y_i)$, $i=1,\ldots,n$. The estimator $\tilde r_b$ depends on the bandwidth $b$ that is generally different from the bandwidth $h$ used in $\hat r_h$. Moreover, $\tilde r_b$ is based on the kernel $H$ that may differ from the kernel $K$ used in $\hat r_h$. Thus, $K=H$ in the original OSCV implementation of~\citet{HartYi}, but $K\neq H$ in the fully robust OSCV method of~\citet{Savchuk:OSCVnonsmooth}. In the asymptotic sense, computing $\tilde r_b$ by using the data on only one side of an estimation point is equivalent to using all data points in the local linear estimator based on the so-called one-sided kernel $L$ related to the kernel $H$ in the following way:
\begin{equation}
\label{eq:L}
L(u)=2H(u)\frac{S_2-uS_1}{S_2-2S_1^2}I_{[0,\infty)}(u),
\end{equation}
where
\[
S_i=\int_0^\infty u^iH(u)\,du,\quad i=1,2,
\]
and $I_A$ is the indicator function of set $A$.

The one-sided estimator $\tilde r_b$ is used to compute the one-sided cross-validation function defined as
\begin{equation}
\label{eq:OSCVfunction}
\mbox{OSCV}(b)=\frac{1}{n-m}\sum_{i=m+1}^n(\tilde r_b^i(x_i)-Y_i)^2,
\end{equation}
where $\tilde r_b^i$ is the leave-one-out version of $\tilde r_b$. Thus, $\tilde r_b^i(x_i)$ is computed from the observations $(x_1,Y_1),\ldots,(x_{i-1},Y_{i-1})$. The quantity $m$ is the number of the data points that are used to compute $\tilde r_b^{m+1}(x_{m+1})$. It is common to take $m=4$. Let $\hat b_{OSCV}$ denote the minimizer of the OSCV function.

The OSCV function~\eqref{eq:OSCVfunction} is defined by analogy with the cross-validation function of Stone (1977) that is given by
\[
\mbox{CV}(h)=\frac{1}{n}\sum_{i=1}^n\left(\hat r_h^{-i}(x_i)-Y_i\right)^2.
\]
In the above expression $\hat r_h^{-i}$ is the leave-one-out estimator that is the local linear estimator computed from all data except for the $i$th observation. Let $\hat h_{CV}$ denote the minimizer of the CV function.

Let $\mbox{AMASE}_L$ denote the AMASE function for the local linear estimator based on the kernel $L$ in the case when $r$ is smooth. An expression for $\mbox{AMASE}_L$ is obtained from~\eqref{eq:AMASE_smooth} by  everywhere replacing $K$ by $L$. The minimizer of $\mbox{AMASE}_L$ is denoted by $b_n$. It appears that
\begin{equation}
\label{eq:C}
\frac{h_n}{b_n}=\left(\frac{R_K}{\mu_{2K}^2}\cdot\frac{\mu_{2L}^2}{R_L}\right)^{1/5}=\frac{J_K}{J_L}\equiv C,
\end{equation}
where $J_K$ and $J_L$ are computed for the kernels $K$ and $L$, respectively, according to~\eqref{eq:J}. The constant $C$ is referred to as the {\it smooth constant} and is completely determined by the kernels $K$ and $H$.

\citet{HartYi} argued that in the case when $r$ is smooth, the OSCV function is approximately unbiased estimator of $\mbox{MASE}_L+\sigma^2$. This implies that $\hat b_{OSCV}$ estimates $b_n$. These considerations justify the following OSCV method's bandwidth selection rule:
\begin{equation}
\label{eq:h_OSCV}
\hat h_{OSCV}=C\hat b_{OSCV},
\end{equation}
where $C$ is defined by~\eqref{eq:C}. It appears that $\hat h_{OSCV}$ is a consistent estimator of $h_0$ in the case when $r$ is smooth. The OSCV regression estimate is computed as the local linear estimate~\eqref{eq:LLE} based on the bandwidth $\hat h_{OSCV}$.

\citet{Savchuk:OSCVnonsmooth} developed the OSCV theory in the case when the regression function $r$ is nonsmooth. Given that the derivative of $r$ has jumps at the points $\{x^{(t)}\}$, $t=1,\ldots,k$, the asymptotic MASE expansion for the local linear estimator $\hat r_h$ based on the kernel $K$ has the following form:
\begin{equation*}
\mbox{MASE}_K(h)=\mbox{AMASE}_K^*(h)+O\left(h^4+\frac{1}{n^2h^3}\right) +o\left(\frac{1}{nh}\right),
\end{equation*}
where
\begin{equation}
\label{eq:AMASE_nonsmooth}
\mbox{AMASE}_K^*(h)=\frac{R_K\sigma^2}{nh}+h^3B_K\sum_{t=1}^k
f(x^{(t)})\left(r^\prime(x^{(t)}+)-r^\prime(x^{(t)}-)\right)^2.
\end{equation}
The value of $B_K$ is computed for the kernel $K$ according to~\eqref{eq:B}. \citet{Savchuk:OSCVnonsmooth} derived the result~\eqref{eq:AMASE_nonsmooth} in the case of a regression function defined on $[0,1]$, but it also holds in the case of a regression function defined on $[0,a]$. The minimizer of $\mbox{AMASE}_K^*$ has the following form:
\[
h_n^*=\left(\frac{R_K\sigma^2}{3B_K\sum_{t=1}^k f(x^{(t)})(r^\prime(x^{(t)}+)-r^\prime(x^{(t)}-))^2}\right)^{1/4}n^{-1/4}.
\]

Let $\mbox{AMASE}_L^*$ denote the AMASE function that is computed for the local linear estimator based on the kernel $L$ in the case when $r$ is nonsmooth. An expression for $\mbox{AMASE}_L^*$ follows from~\eqref{eq:AMASE_nonsmooth} by everywhere replacing $K$ by $L$. Let $b_n^*$ denote the minimizer of $\mbox{AMASE}_L^*$. It follows that
\begin{equation}
\label{eq:Cstar}
\frac{h_n^*}{b_n^*}=\left(\frac{R_K}{B_K}\cdot
\frac{B_L}{R_L}\right)^{1/4}\equiv C^*,
\end{equation}
where $R_K,\ B_K$ and $R_L,\ B_L$ are computed for the kernels $K$ and $L$, respectively, according to\eqref{eq:R_mu_notation} and~\eqref{eq:B}. The constant $C^*$ is completely determined by the kernels $K$ and $H$ and is referred to as the {\it nonsmooth constant}.

In the nonsmooth case, the relative bandwidth bias increase due to inappropriate using $C$ instead of $C^*$ can be assessed as
\begin{equation}
\label{eq:Cmismatch}
E_C=\frac{C\hat b_{OSCV}-C^*\hat b_{OSCV}}{C^*\hat b_{OSCV}}\cdot100\%=\frac{C-C^*}{C^*}\cdot100\%.
\end{equation}
It follows from the results of~\citet{Savchuk:OSCVnonsmooth}, that $E_C<7.01\%$ for such frequently used kernels as Epanechnikov and quartic. However, in the case of the Gaussian kernel $\phi$, defined as $\phi(u)=(2\pi)^{-1/2}\exp(-u^2/2)$, the discrepancy $E_C=16.74\%$. Indeed in the Gaussian case the smooth constant $C_\phi=0.6168$, and the nonsmooth constant $C_\phi^*=0.5284$. Simulation study of~\citet{Savchuk:OSCVnonsmooth} confirms that OSCV tends to produce too large bandwidths in the case when $r$ is nonsmooth and $K=H=\phi$. Our experience with smoothing suggests that the relative bias of 7\% has a negligible effect on performance of an estimator. However, the bias increase of 16\% indicates that the methods requires a bias correction.

Theoretically, the bandwidth bias in the case when $r$ is nonsmooth can be eliminated by replacing $C$ by $C^*$ in the OSCV bandwidth rule~\eqref{eq:h_OSCV}. However, such replacement should be justified by either prior information about nonsmoothness of $r$ or the existence of cusps in $r$ should be evident from the scatter diagram of the data. The cusps may be masked by the noise in the data, so the analyst may erroneously apply a smooth version of the OSCV rule~\eqref{eq:h_OSCV} to a nonsmooth function $r$.

Interestingly, the bandwidth bias introduced by inappropriate use of the smooth constant $C$ in the nonsmooth case has a trivial effect on MASE, at least asymptotically. This is shown in~\citet{Savchuk:OSCVnonsmooth} based on the following measure of error:
\[
E_{MASE}=\left(\frac{\mbox{AMASE}_K^*(Cb_n^*)}{\mbox{AMASE}_K^*(h_n^*)}-1\right)\cdot 100\%.
\]
The bandwidth $h_n^*$ is an asymptotic analog of the bandwidth $C^*\hat b_{OSCV}$, whereas the quantity $Cb_n^*$ is asymptotically equivalent to the OSCV bandwidth~\eqref{eq:h_OSCV} that is not justified for the case when $r$ is nonsmooth. It appears that
\[
E_{MASE}=\left(\frac{3}{4}x+\frac{1}{4x^3}-1\right)\cdot 100\%,
\]
where
\[
x=\left(\frac{B_L}{B_K}\right)^{1/4}\left(\frac{R_K}{R_L}\right)^{1/20}\left(\frac{\mu_{2K}^2}{\mu_{2L}^2}\right)^{1/5}.
\]
The quantity $E_{MASE}$ is 0.72\% for the Epanechnikov kernel, it is 0.73\% for the quartic kernel, but it is equal to 4.02\% for the Gaussian kernel. Attempts to improve statistical properties of the OSCV method in the case when $r$ is nonsmooth and $K=\phi$, resulted in the {\it fully robust OSCV} method proposed by~\citet{Savchuk:OSCVnonsmooth}.

\section{Fully Robust OSCV}

In the fully robust OSCV method one uses the fact that the rescaling constants $C$ and $C^*$ are completely determined by the kernels $K$ and $H$. For fixed $K$ one may choose $H$ such that $C=C^*$. A kernel $H$ that produces such equality is called {\it robust} since it makes the OSCV method consistent regardless of smoothness of $r$. The $E_{MASE}$ measure for a robust kernel is identical zero.

\citet{Savchuk:OSCVnonsmooth} fixed $K=\phi$ and found a robust kernel in the following family:
\begin{equation}
\label{eq:H_I}
H_I(x)=(1+\alpha)\phi(x)-\frac{\alpha}{\sigma}\phi\left(\frac{x}{\sigma}\right), \quad x,\alpha\in\mathbb{R},\ \sigma>0.
\end{equation}
The subscript ``$I$'' is used to indicate that the kernels~\eqref{eq:H_I} originate from the indirect cross-validation method of~\citet{SavchukHartSheather:ICV}. The robust kernel used in~\citet{Savchuk:OSCVnonsmooth} has
\begin{equation}
\label{eq:H_I_parameters}
\begin{array}{l}
\displaystyle{\alpha=0.0000879985198548436}\ \ \mbox{and}\\
\displaystyle{\sigma=10.}
\end{array}
\end{equation}
The solution~\eqref{eq:H_I_parameters} was originally found in the way explained below.

\begin{figure}
\begin{center}
\begin{tabular}{ccc}
\epsfig{file=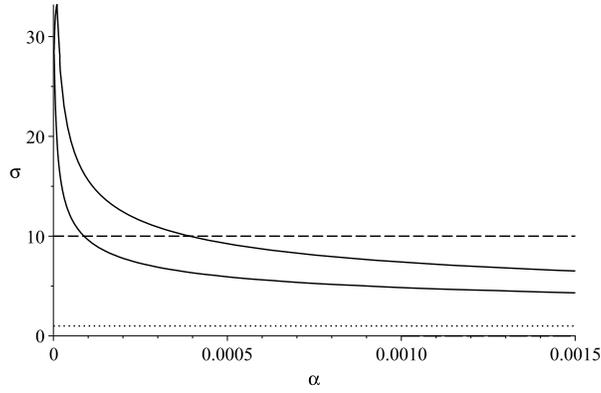,height=150pt}
\end{tabular}
\vspace{-0.5cm} \caption{Robust negative-tailed kernels. The dotted and dashed lines correspond to $\sigma=1$, and $\sigma=10$, respectively. \label{fig:H_Irobust}}
\end{center}
\end{figure}

The family~\eqref{eq:H_I} produces probability density functions for $-1\leq\alpha\leq 0$ and $\sigma>0$ or for $\alpha>0$ and $\alpha/(1+\alpha)\leq \sigma\leq 1$. We did not find any nonnegative robust kernels in the range $-1\leq \alpha\leq 100$. We, thus, started to search for robust kernels in the region $\alpha>0$ and $\sigma>1$ that corresponds to the kernels with negative tails. Observe that $\alpha=0$ yields $H_I\equiv\phi$. Since the constants $C_\phi$ and $C_\phi^*$ are close, we looked for insignificant modification of $\phi$ that correspond to the values of $\alpha$ close to zero. Figure~\ref{fig:H_Irobust} shows the robust negative-tailed kernels in the range $0.00001\leq \alpha\leq 0.0015$.

All kernels in Figure~\ref{fig:H_Irobust} are close to $\phi$. \citet{Savchuk:OSCVnonsmooth} arbitrary selected $\sigma=10$ and used the solution~\eqref{eq:H_I_parameters}. The other solution at $\sigma=10$ has $\alpha=0.0003912884532000514$. The corresponding kernel has somewhat larger $L^2$ distance compared to the kernel defined by~\eqref{eq:H_I_parameters}, but still performs almost identical to it. Actually, all robust kernels shown in Figure~\ref{fig:H_Irobust} are really close and perform similarly. In what follows, we concentrate on using $H_I$ with the values of the parameters as in~\eqref{eq:H_I_parameters}.

The kernel $H_I$ has the unique rescaling constant $C_{I}=0.5217$ that is appropriate in both smooth and nonsmooth cases. Let $\hat b_I$ denote the minimizer of the OSCV function~\eqref{eq:OSCVfunction} that is computed based on $H_I$. The corresponding bandwidth that is used to compute a regression estimate is $\hat h_I=C_I\hat b_I$. In what follows, $\hat b_{OSCV}$ corresponds to the minimizer of the OSCV curve based on the Gaussian kernel $\phi$, and $\hat h_{OSCV}=C_\phi\hat b_{OSCV}$.

The kernels $\phi$ and $H_I$ look virtually the same when plotted on the interval $[-4,4]$. It turns out that $H_I(x)<0$ for $|x|>4.85$. It is also remarkable that the tails of $H_I(x)$ are close to zero even for ``large'' $x$. Thus, $H_I(\pm10)=-2.13\cdot10^{-6}$. 
It appears that the efficiency of $H_I$, computed according to~\citet{Wand:KS}, is 0.9552, that is even larger than 0.9512 in the case of $\phi$. Nevertheless, we only use $H_I$ for the cross-validation purposes since it has negative tails.

\begin{figure}[h]
\begin{center}
\epsfig{file=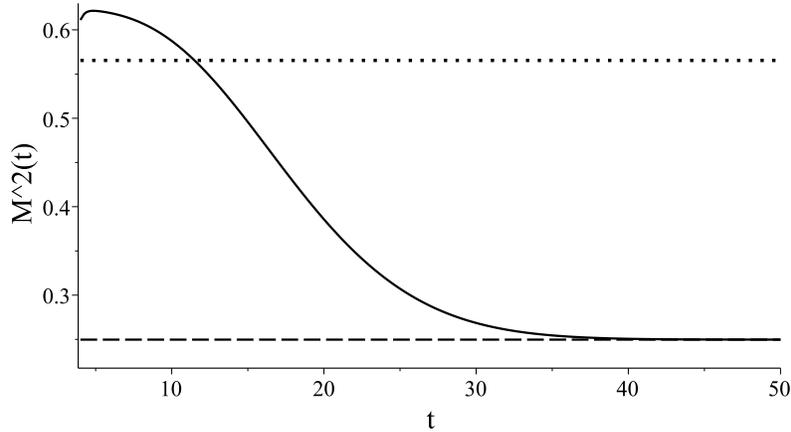,height=170pt}
\vspace{-0.25cm}
\caption{$M_{L_I}^2(t)$ for $4\leq t\leq50$. Dashed line shows $\mu_{2L_I}^2=0.2497$; dotted line shows $\mu_{2L_\phi}^2=0.5654$.\label{fig:mu2L_I}}
\end{center}
\end{figure}

Let $L_\phi$ and $L_I$ denote the one-sided counterparts of $\phi$ and $H_I$, respectively, computed according to~\eqref{eq:L}.  Closeness of $L_{\phi}$ and $L_I$ on the interval $[0,4]$ even seems to contradict the fact that $C_\phi=0.6168$, whereas $C_I=0.5217$. It follows from~\eqref{eq:C} that the discrepancy in $C_\phi$ and $C_I$ is caused by the difference in the constants $J_{L_\phi}$ and $J_{L_I}$, obtained from $L_\phi$ and $L_{I}$ according to~\eqref{eq:J}. Indeed, $J_{L_\phi}=1.2586$, whereas $J_{L_I}=1.4882$. Consider the numerical values of the constituents of $J_{L_\phi}$ and $J_{L_I}$:
\[
\begin{array}{ll}
R_{L_\phi}=1.7860,&R_{L_I}=1.8230,\\
\mu_{2L_\phi}^2=0.5654,&\mu_{2L_I}^2=0.2497.\\
\end{array}
\]
The values $R_{L_\phi}$ and $R_{L_I}$ are quite close. It appears that the squared second moment is the culprit in causing the discrepancy between $J_{L_\phi}$ and $J_{L_I}$. The mismatch in $\mu_{2L_\phi}^2$ and $\mu_{2L_I}^2$ must be explained by different behaviour of $L_\phi$ and $L_I$ in the tails.
Define the following integrals for a one-sided kernel $L$:
\[
\begin{array}{l}
\displaystyle{M_L(t)=\int_0^t u^2L(u)\,du,}\\\\
\displaystyle{F_L(t)=\left(\frac{\int_0^t L(u)^2\,du}{M_L^2(t)}\right)^{1/5}}.
\end{array}
\]
Observe that $M_L(t)\rightarrow\mu_{2L}$ and $F_L(t)\rightarrow J_L$ as $t\rightarrow\infty$.
Figure~\ref{fig:mu2L_I} contains a plot of $M_{L_I}^2(t)$ for $4\leq t\leq 50$. The dashed and the dotted horizontal lines indicate the values of $\mu_{2L_I}^2$ and $\mu_{2L_\phi}^2$, respectively. The curve $M_{L_\phi}^2(t)$ is not shown since it is indistinguishable from the dotted line for $t\geq 5$. The figure shows that $M_{L_I}(t)$ substantially deviates from $\mu_{2L_\phi}^2$ for $t$ larger than about 15.

\begin{figure}[h]
\begin{center}
\epsfig{file=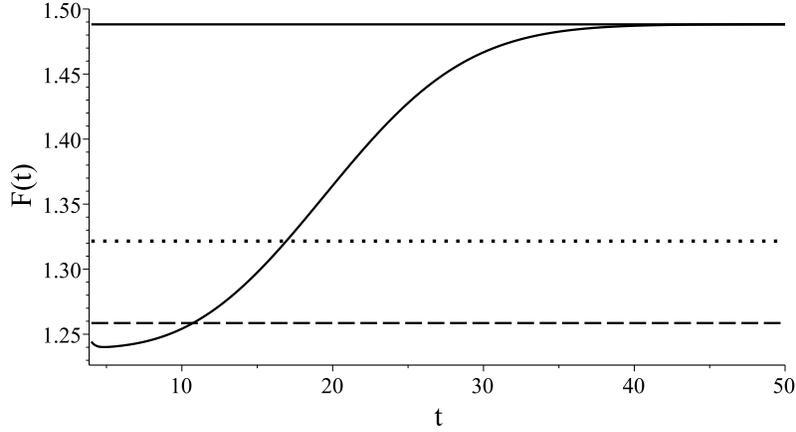,height=170pt}
\vspace{-0.25cm}
\caption{$F_{L_I}(t)$ for $4\leq t\leq50$. Dashed line shows $J_{L_\phi}=1.2586$; solid horizontal line shows $J_{L_I}=1.4882$; dotted line shows $1.05J_{L_\phi}$.\label{fig:F_LI}}
\end{center}
\end{figure}

Figure~\ref{fig:F_LI} shows a plot of $F_{L_I}$ for $4\leq t\leq 50$. The graph of $F_{L_\phi}$ is not shown since for $t\geq 5$ it practically coincides with the dashed line showing $J_{L_\phi}$.
It turns out that $F_{L_I}(t)\geq 1.05J_{L_\phi}$ for $t\geq16.92$. This suggests that as long as $L_{I}$ is not evaluated at a value larger than about 16.92, there is no reason to think that, in practical sense, using $L_I$ is any different than using $L_\phi$. It follows from~\eqref{eq:w}, \eqref{eq:t}, and~\eqref{eq:OSCVfunction} that we evaluate $L_I$ at values of the form $(x_i-x_j)/b$, where $b$ is a bandwidth, and $x_1<x_2<\ldots<x_n$ are the design points on the interval $[0,a]$. Observe that $\displaystyle{\max_{i,j}|x_i-x_j|\approx a}$.
In the case when $r$ is smooth, the $\mbox{AMASE}_{L_I}$-optimal bandwidth, that is a proxy to the OSCV minimizer $\hat b_I$, is less than $a/16.92$ for
\begin{equation}
\label{eq:n_smooth}
n>\left(\frac{16.92}{a}\right)^5\frac{R_{L_I}\sigma^2}{\mu_{2L_I}^2\int_0^a\left(r\,^{\prime\prime}(x)\right)^2f(x)\,dx}=
1.0124\cdot 10^7\frac{\sigma^2}{a^5\int_0^a\left(r\,^{\prime\prime}(x)\right)^2f(x)\,dx}.
\end{equation}
In the nonsmooth case, the $\mbox{AMASE}_{L_I}^*$-optimal bandwidth is less than $a/16.92$ given that
\begin{multline}
\label{eq:n_nonsmooth}
n>\left(\frac{16.92}{a}\right)^4\frac{R_{L_I}\sigma^2}{3B_{L_I}\sum_{t=1}^k
f(x^t)\left(r^\prime(x^t+)-r^\prime(x^t-)\right)^2}=\\
9.7408\cdot10^5\frac{\sigma^2}{a^4\sum_{t=1}^k
f(x^t)\left(r^\prime(x^t+)-r^\prime(x^t-)\right)^2}.
\end{multline}

In~\eqref{eq:n_smooth} and~\eqref{eq:n_nonsmooth}, the sample size is proportional to $\sigma^2$. This suggests that larger $n$ is required for noisier data to make the difference between using $L_\phi$ and $L_I$ essential.

For numerical illustration of~\eqref{eq:n_smooth} and~\eqref{eq:n_nonsmooth}, we use three regression functions, $r_1$, $r_2$, and $r_3$, that originate from the simulation study of~\citet{Savchuk:OSCVnonsmooth} and are defined in the Appendix.
Figure~\ref{fig:fun_data} shows the graphs of $r_1$, $r_2$, and $r_3$ along with the typical data sets generated for specified $n$ and $\sigma$.
\begin{figure}
\begin{center}
\begin{tabular}{ccc}
$n=100$, $\sigma=1/1000$&$n=300$, $\sigma=1/500$&$n=1000$, $\sigma=1/250$\\
\epsfig{file=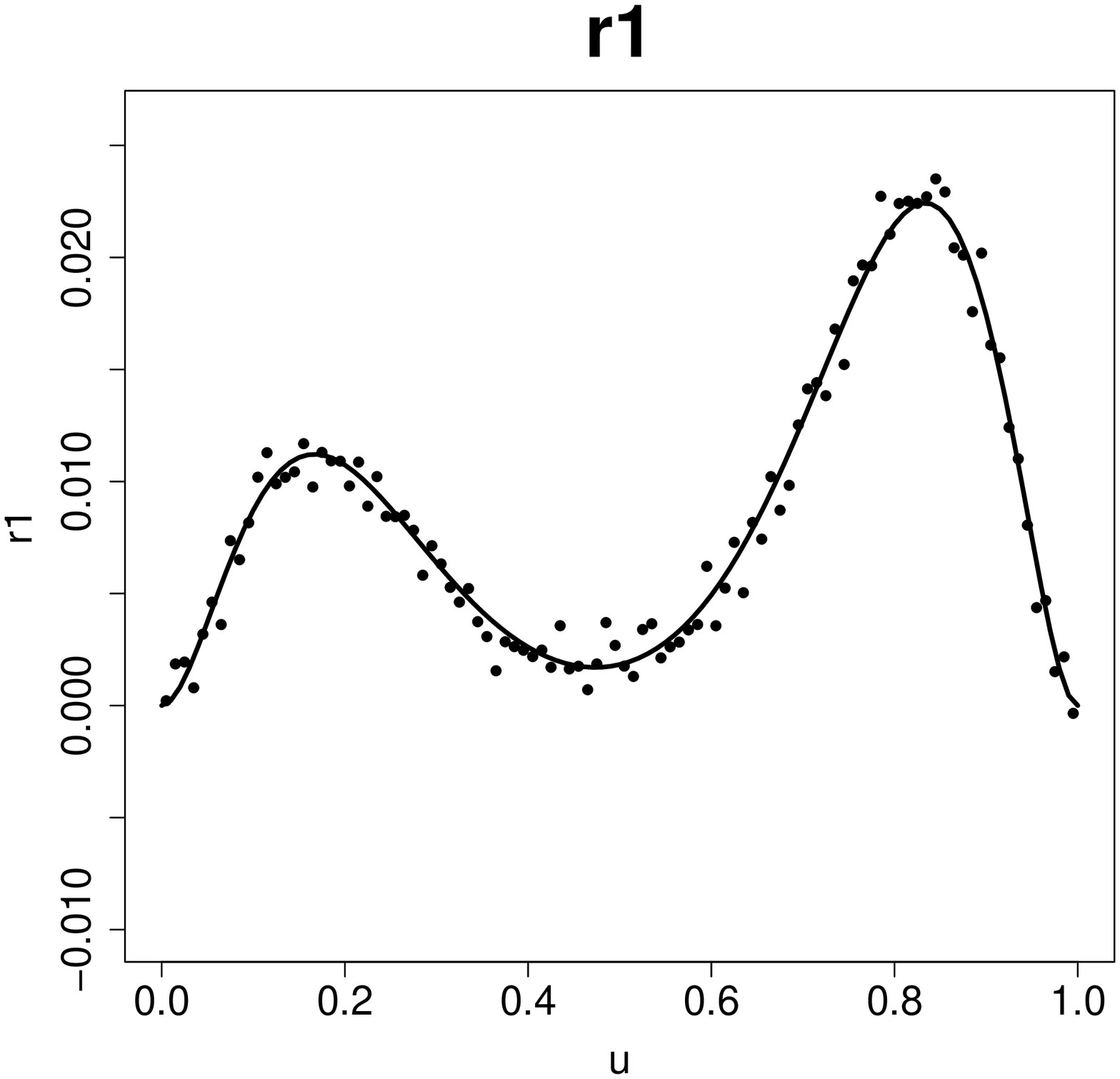,height=150pt}&\hspace{-0.25cm}\epsfig{file=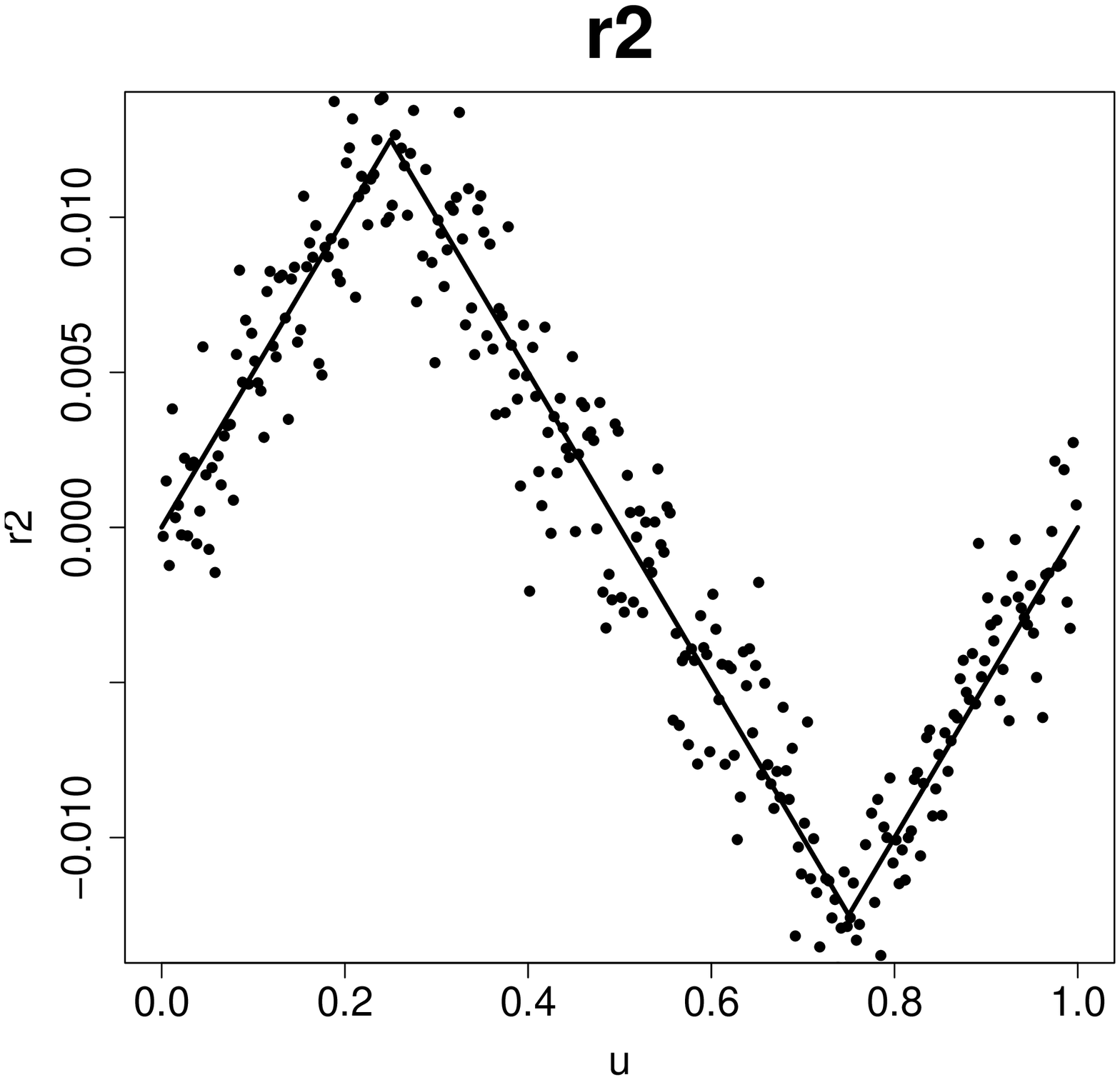,height=150pt}&\hspace{-0.25cm}\epsfig{file=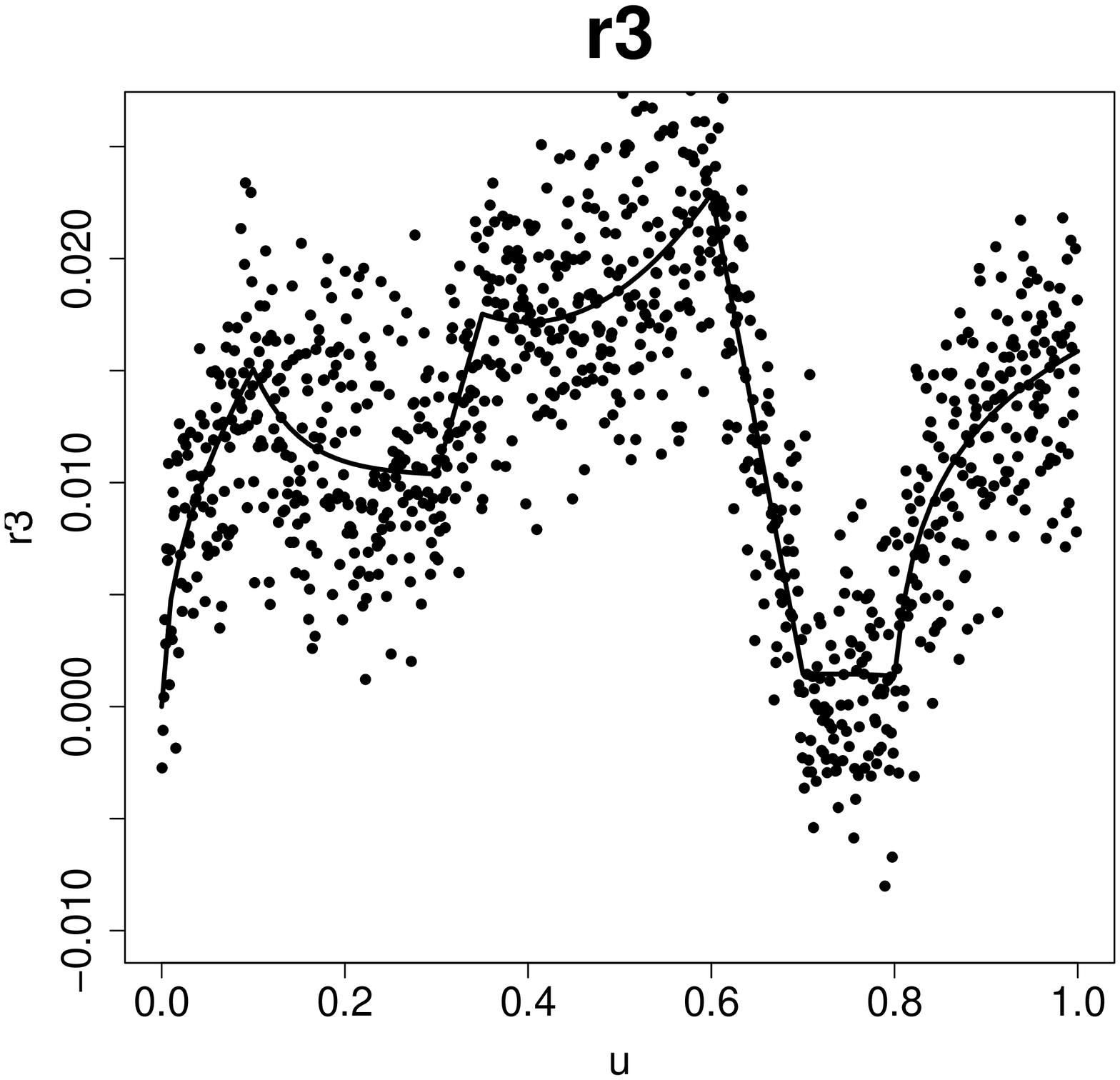,height=150pt}
\end{tabular}
\vspace{-0.5cm} \caption{Regression functions $r_1$, $r_2$, $r_3$ and generated data. \label{fig:fun_data}}
\end{center}
\end{figure}

Table~\ref{tab:n} shows   the smallest $n$ that satisfies~\eqref{eq:n_smooth} for $r_1$ and~\eqref{eq:n_nonsmooth} for $r_2$ and $r_3$ in the case $\sigma=1/500$ and $f(x)=1$.
\begin{table}[h]
\begin{center}
\caption{Smallest $n$ at which the difference between $\mu_{L_\phi}^2$ and $\mu_{2L_I}^2$ becomes essential in the case $f=1$, and $\sigma=1/500$.\label{tab:n}}
\begin{tabular}{|l|c|c|c|}
\hline
Function&$r_1$&$r_2$&$r_3$\\
\hline $n$&17&195&10\\
\hline
\end{tabular}
\end{center}
\end{table}
Since $r_3$ is the least smooth function of the three, it requires using smaller bandwidths and, consequently, involves the tail of $L_I$ for smaller $n$ compared to the other two functions. The results in Table~\ref{tab:n} indicate that the difference between using $L_I$ and $L_{\phi}$ might be evident in finite samples.

\begin{figure}
\begin{center}
\epsfig{file=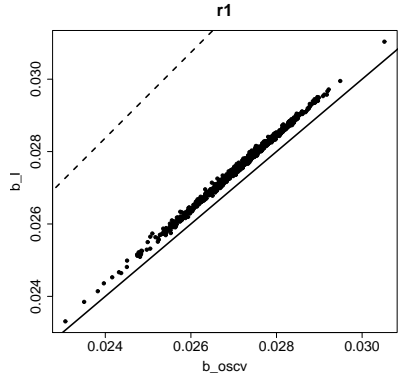,height=150pt}
\vspace{-0.5cm} \caption{Scatter plots of $\hat b_I$ versus $\hat b_{OSCV}$ in the case of $r_1$, $n=1000$, and $\sigma=1/500$.  \label{fig:bI_boscv}}
\end{center}
\end{figure}

The results of the numerical study of~\citet{Savchuk:OSCVnonsmooth} are used to compare the finite sample performances of the $H_I$-and $\phi$-based OSCV versions. The sample sizes considered in~\citet{Savchuk:OSCVnonsmooth} are $n=50$, 100, 300, and 1000, and the Gaussian noise levels are $\sigma=1/250$, 1/500, and 1/1000. For a random variable $Y$ defined in each replication of a simulation, let $\hat E(Y)$, $\hat{SD}(Y)$ and $\hat M(Y)$ denote the average, standard deviation, and median of $Y$ over 1000 replications with $r$, $n$, and $\sigma$ being fixed. One of the most important observations in the numerical study of~\citet{Savchuk:OSCVnonsmooth} is that $\hat E(\hat h_I/\hat h_{OSCV})\approx\hat{SD}(\hat h_I/\hat h_{OSCV})\approx C_I/C_{\phi}=0.85$ for all considered regression functions, noise levels and sample sizes. This implies that $\hat b_I\approx\hat b_{OSCV}$. This result is not surprising in the nonsmooth case, where it is expected that for ``large'' $n$
\[
\displaystyle{\hat b_I\approx\frac{C_\phi^*}{C_I}\hat b_{OSCV}=1.0128\cdot\hat b_{OSCV}}.
\]
The corresponding large sample result in the smooth case is
\[
\displaystyle{\hat b_I\approx\frac{C_\phi}{C_I}\hat b_{OSCV}=1.1823\cdot\hat b_{OSCV}}.
\]
Thus, in the smooth case $\hat b_I$ is expected to be somewhat larger than $\hat b_{OSCV}$ for ``large'' $n$. Figure~\ref{fig:bI_boscv} shows the scatter plots of $\hat b_{I}$ versus $\hat b_{OSCV}$ in the case of $r_1$, $n=1000$, and $\sigma=1/500$. The solid line in the plot shows the 45 degrees line that passes through the origin. The dashed line passes through the origin and has the slope equal to 1.1823.
The points on the graph form a line that lies between the solid and dashed lines, but substantially closer to the former one compared to the latter one. Larger sample size might be needed for the points to lie closer to the line with the slope of 1.1823.

The result $\hat b_I\approx\hat b_{OSCV}$ is a consequence of the fact that the $H_I$- and $\phi$-based OSCV curves computed for the same data set are usually drastically close except in the neighborhood of zero, where the $H_I$-based curve might occasionally exhibit spurious bumps, as illustrated in the data examples in Section~\ref{sec:Examples}.

Since $C_{\phi}>C_I$, the result $\hat b_I\approx b_{OSCV}$ implies that the $H_I$-based OSCV version is expected to produce too low bandwidths in the smooth case. For assessment of the finite sample relative bandwidth bias by the $H_I$- and $\phi$-based OSCV versions, we use the numerical data of~\citet{Savchuk:OSCVnonsmooth} to compute
\[
\Delta_B=\frac{\hat M(\hat h)-\hat M(\hat h_0)}{\hat M(\hat h_0)}\cdot100\%.
\]
Table~\ref{tab:Delta_B} contains the values of $\Delta_B$ in the cases $\hat h=\hat h_{I}$ and $\hat h_{OSCV}$ for all regression functions, $n=100$, 300, 1000 and $\sigma=1/500$.
\begin{table}[h]
\begin{center}
\caption{ Values of $\Delta _B$ for $r_1$, $r_2$, and $r_3$ in the case $\sigma=1/500$ and $n=100$, 300, and 1000.\label{tab:Delta_B}}
\begin{tabular}{|l|c|c|c||c|c|c|}
\hline
Method&\multicolumn{3}{|c||}{$H_I$-based OSCV}&\multicolumn{3}{|c|}{$\phi$-based OSCV}\\
\hline
Function&$r_1$&$r_2$&$r_3$&$r_1$&$r_2$&$r_3$\\
\hline
$n=100$&-12.97&-8.33&-3.37&2.10&8.00&13.46\\
\hline
$n=300$&-12.76&-9.45&-5.33&1.91&6.48&11.40\\
\hline
$n=1000$&-12.77&-8.84&-5.65&1.48&6.87&10.87\\
\hline
\end{tabular}
\end{center}
\end{table}
In the case of $r_1$, the $\phi$-based OSCV method produces bandwidths that are slightly biased upward, whereas the $H_I$-based OSCV version has the relative bandwidth bias of about -13\%. In the case of $r_3$, the value of $\Delta_B$ for $H_I$-based OSCV is still negative but much closer to zero, whereas the $\phi$-based OSCV method has $\Delta_B>10\%$ for all considered sample sizes. The case of $r_2$ is intermediate: the values of $\Delta_B$ for ordinary OSCV and $H_I$-based OSCV are similar in magnitude but have opposite signs. Both versions of the OSCV method seem to be tricked by the function $r_2$ that has two cusps that can be easily masked by the data's noise, as it is illustrated in Figure~\ref{fig:fun_data}.

The measure $\Delta_B$ can be thought of an empirical analog of $E_C$ in the nonsmooth case. Observe that for the $\phi$-based OSCV method, the values of $\Delta_B$ in Table~\ref{tab:Delta_B} do not approach the theoretical result $E_C=16.73\%$  even in the case of $r_3$ and $n=1000$.

\begin{figure}[h]
\begin{center}
\begin{tabular}{cc}
{\bf(a)}&{\bf(b)}\\
\epsfig{file=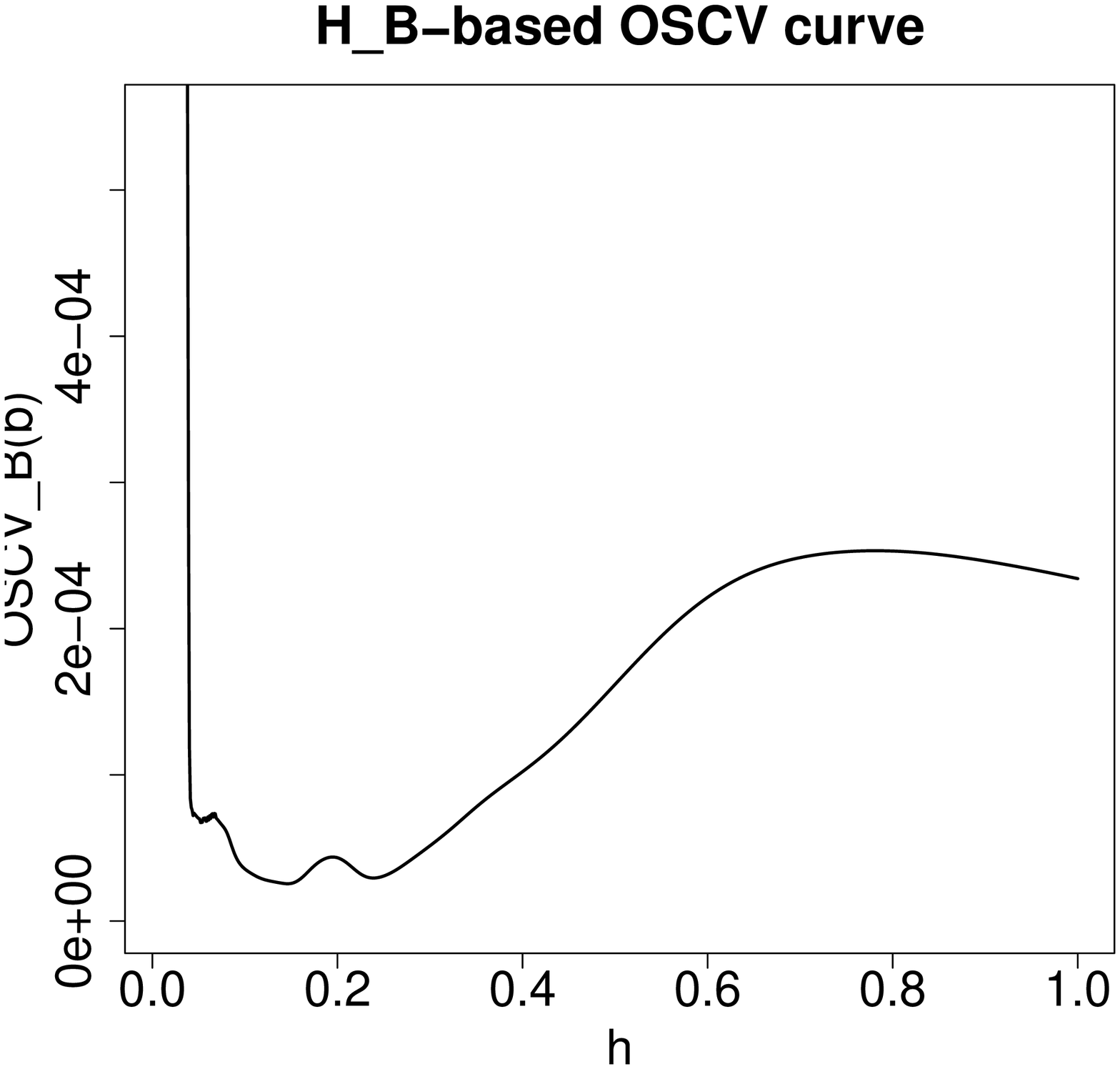,height=180pt}&\epsfig{file=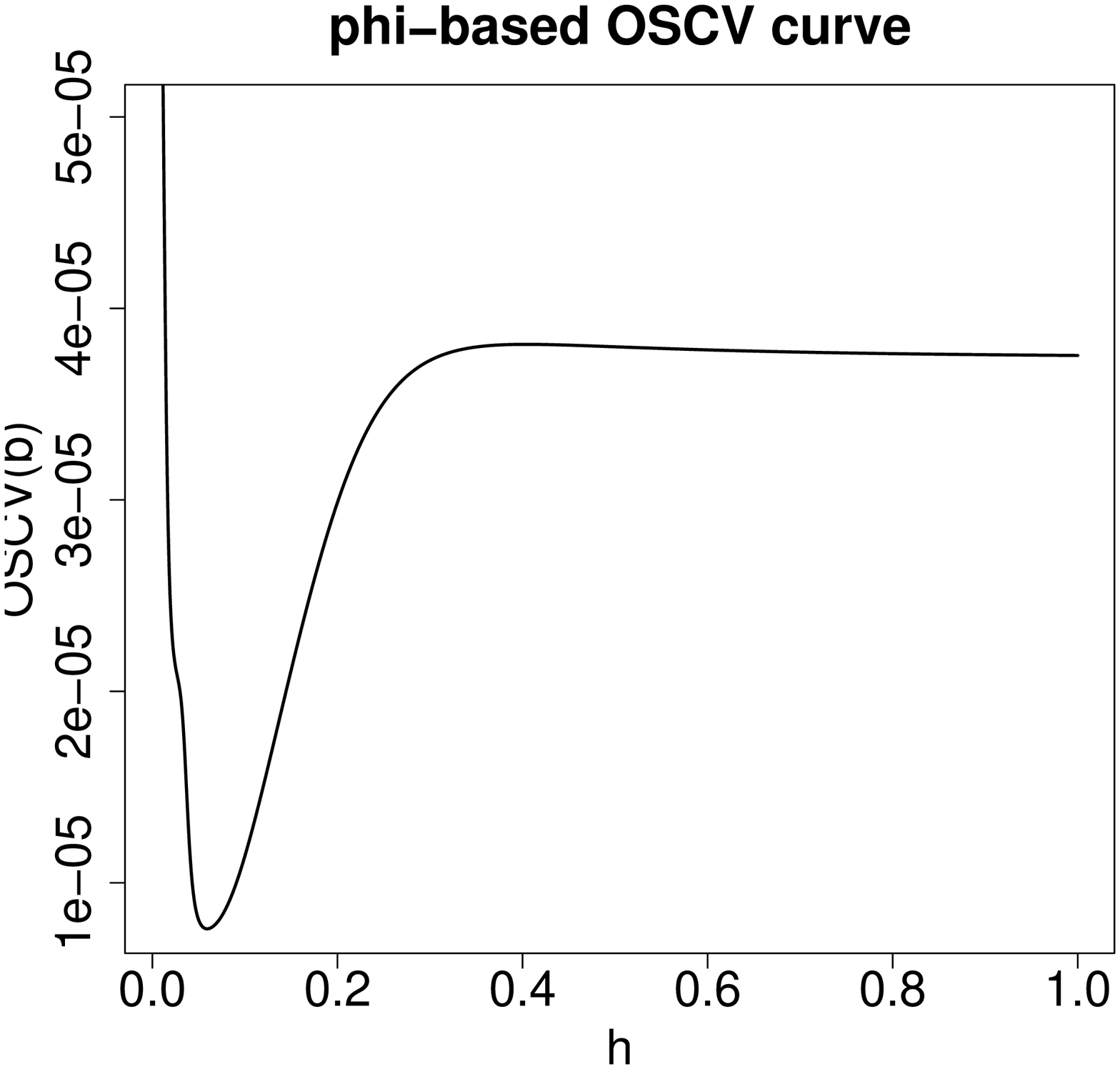,height=180pt}\\
\end{tabular}
\vspace{-0.25cm}
\caption{{\bf(a)} $H_B$- and {\bf(b)} $\phi$-based OSCV curves for the data generated from $r_1$ in the case of $n=100$, $\sigma=1/500$, and the $\mbox{Uniform}(0,1)$ design.\label{fig:bimod_r1_uniform}}
\end{center}
\end{figure}

The wiggles in the $H_I$-based OSCV curve are shown to be caused by negativity of the tails of $H_I$. This inspired a new search of nonnegative robust kernel that resulted in several robust bimodal kernels. One of the kernels, $H_B$, is defined as
\begin{equation}
\label{eq:H_B}
H_B(x)=5\phi(10(x+\mu))+5\phi(10(x-\mu)),
\end{equation}
where $\mu=0.412071682$. The rescaling constant for $H_B$ is $C_B=0.1932$.
We empirically found that bimodality of $H_B$ is associated with producing OSCV curves with several local minima that are often of comparable sizes. This is illustrated in Figure~\ref{fig:bimod_r1_uniform} {\bf(a)}, that shows a typical $H_B$-based OSCV curve in the case of $r_1$, $\sigma=1/500$, $n=100$ and the $\mbox{Uniform}(0,1)$ design. The corresponding $\phi$-based OSCV curve, shown in Figure~\ref{fig:bimod_r1_uniform} {\bf(b)}, is smooth and has one local minimum.

The other newly found robust bimodal kernels perform similarly to $H_B$. This supports the empirically derived conclusion of \citet{Savchuk:OSCVnonsmooth} that a ``good'' robust kernel should be unimodal and nonnegative. The problem of finding such a kernel is still open.

\section{Data Examples\label{sec:Examples}}

Performances of the $H_I$- and $\phi$-based OSCV versions are further compared on the following two data examples.

\subsection{Example 1 (Fuel consumption).\label{sec:car}}

\begin{figure}[h]
\begin{center}
\begin{tabular}{cc}
{\bf(a)}&{\bf(b)}\\
\epsfig{file=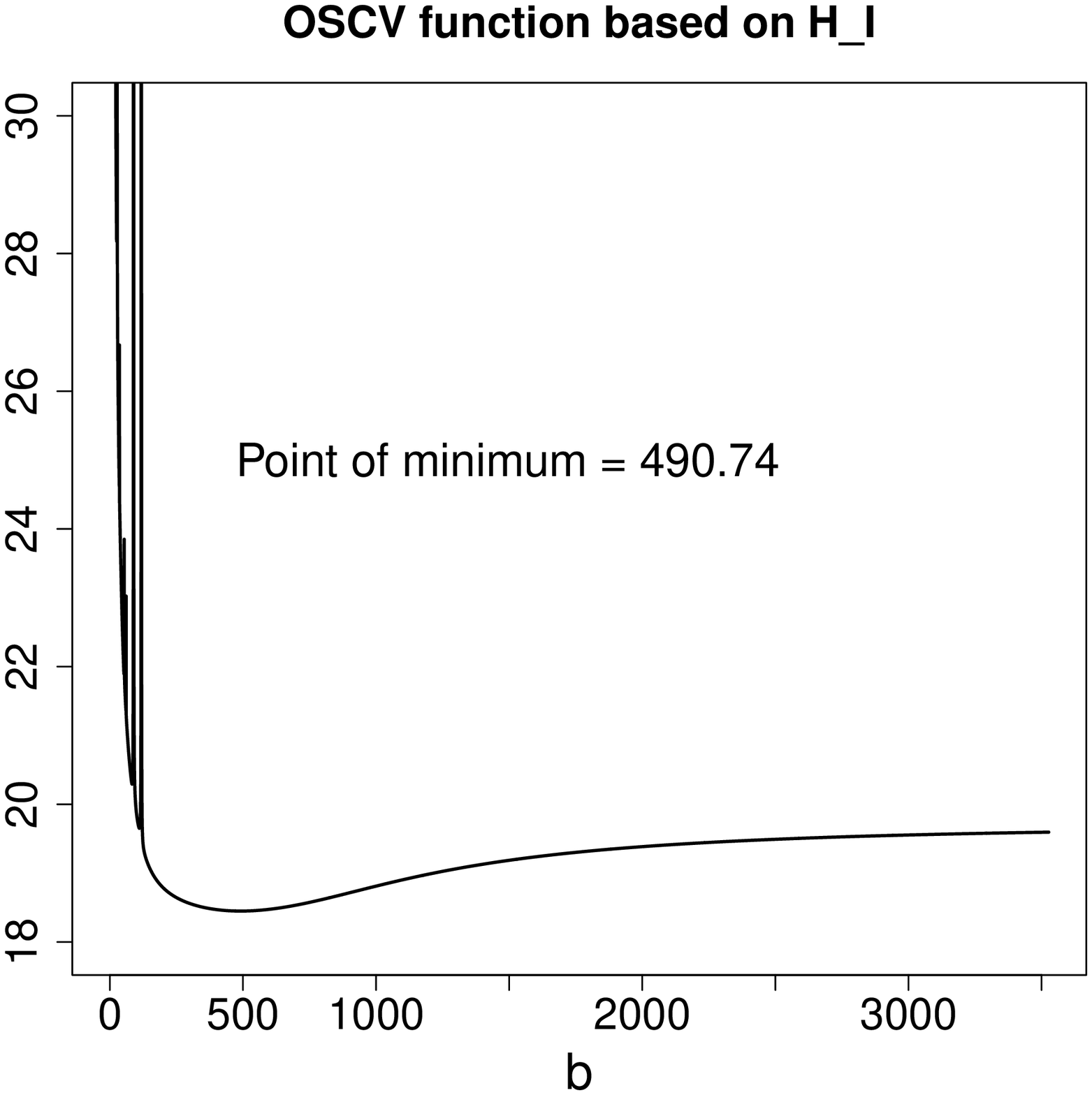,height=180pt}&\epsfig{file=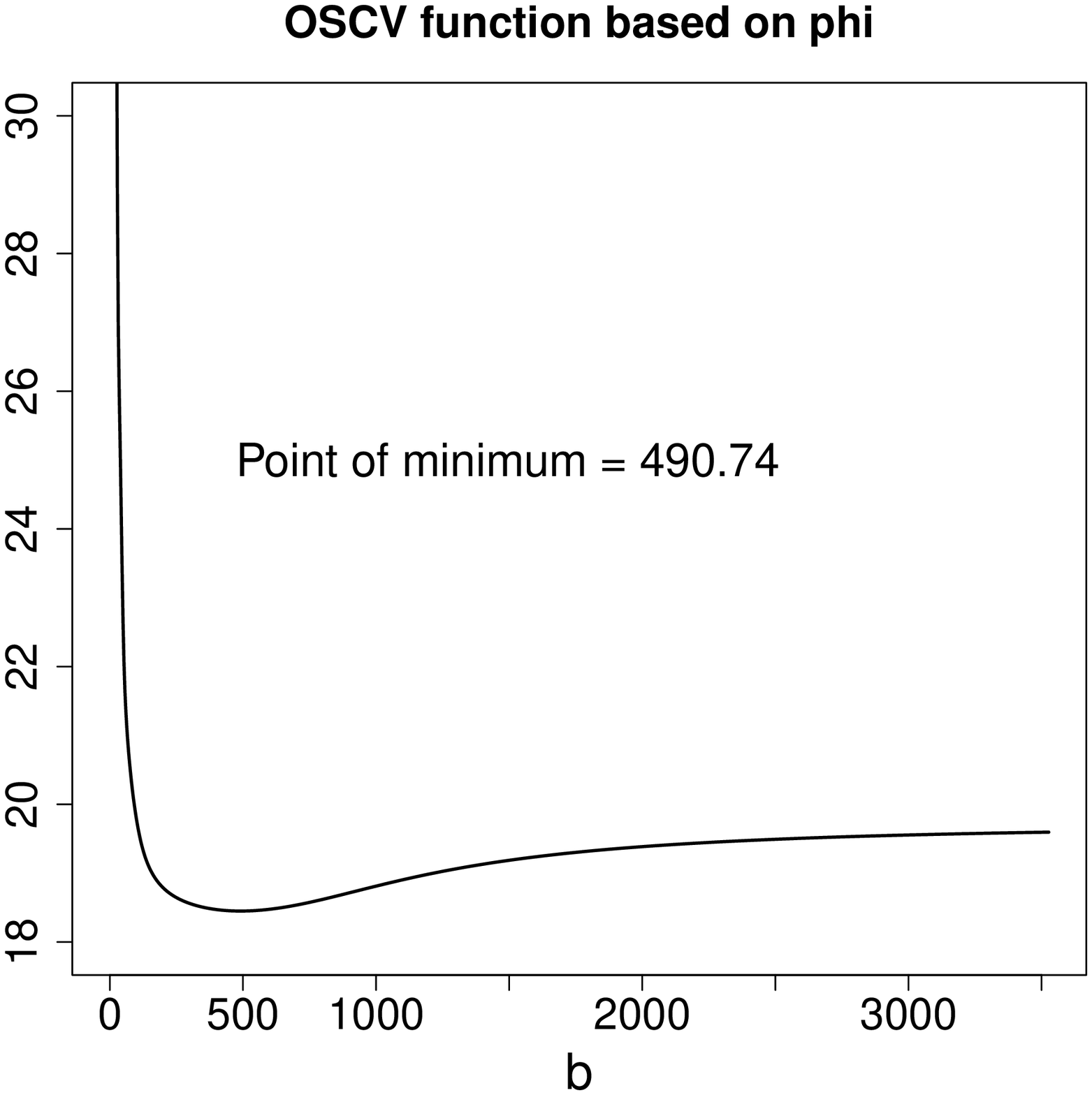,height=180pt}\\
\end{tabular}
\vspace{-0.25cm}
\caption{{\bf(a)} $H_I$- and {\bf(b)} $\phi$-based OSCV curves for the mpg and car weight data.\label{fig:car}}
\end{center}
\end{figure}

\begin{figure}[h]
\begin{center}
\epsfig{file=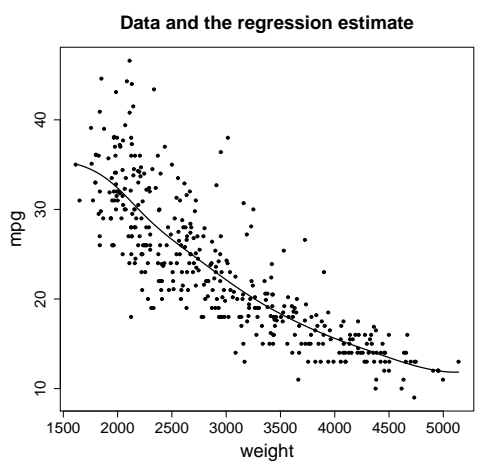,height=180pt}
\caption{Regression estimate based on $\hat h_I=256.02$ for the data on mpg and car weight.\label{fig:LLE_mpg_ROSCV}}
\end{center}
\end{figure}

\noindent The data on car city-cycle fuel consumption in miles per gallon (mpg) can be downloaded from \url{http://archive.ics.uci.edu/ml/datasets/Auto+MPG} with the required citation of~\citet{Lichman:2013}. The same data set is used in~\citet{Savchuk:OSCVnonsmooth}, but in this article we consider dependence of mpg ($y$) on car weight ($x$) instead of horsepower. Let $\delta_i=x_i-x_{i-1}$, $i=2,\ldots,n$. Figure~\ref{fig:car} shows the OSCV curves based on $\phi$ and $H_I$ for $b>\displaystyle{\min_{i}\delta_i}$.
For these data $\hat b_I\approx \hat b_{OSCV}\approx490.74$. The OSCV curves based on $H_I$ and $\phi$ are quite similar except for the values of $b$ near zero, where the OSCV curve based on $H_I$ has spurious wiggles, whereas the OSCV curve based on $\phi$ is smooth. Let $\hat h_{PI}$ denote the Ruppert-Sheather-Wand plug-in bandwidth computed for a given data set. The bandwidths selected by different methods for the data on fuel consumption are shown in the table below.
\begin{center}
\begin{tabular}{|c|c|c|c|}
\hline
$\hat h_{OSCV}$&$\hat h_I$&$\hat h_{CV}$&$\hat h_{PI}$\\
\hline 302.71&256.02&270.64&263.67\\
\hline
\end{tabular}
\end{center}
The local linear regression estimate based on $\hat h_I$ is shown in Figure~\ref{fig:LLE_mpg_ROSCV}. The estimates based on the other bandwidths from the above table are similar.

The $H_I$-based OSCV curve in Figure~\ref{fig:car} {\bf(a)} behaves almost like a discontinuous function for ``small'' $b$.
Alternating sign of $H_I((x_i-x_j)/b)$ for $x_j<x_i$, $i,j=1,\ldots,n$, is one of the factors that occasionally produces very ``small'' sum of weights in the denominator of $\tilde r_b^i(x_i)$ for certain values of $i$ and $b$. The resulting ``large'' value of $\tilde r_b^i(x_i)$ produces ``large'' squared deviation in the OSCV function~\eqref{eq:OSCVfunction} that causes a spike in the OSCV curve at the corresponding value of $b$.

The mpg and car weight example illustrates a typical behaviour of $H_I$-based OSCV for a data set of size $n\geq 100$. Spurious wiggles in the $H_I$-based OSCV curve usually appear for ``small'' $b$ and do not interfere with the problem of determining $\hat b_I$. For $n<100$ the wiggles may occasionally produce a fake global minimum of the $H_I$-based OSCV curve, as it is illustrated by the example in the following section.

\subsection{Example 2 (Weight of rabbits).\label{sec:rabbit}}

\begin{figure}[h]
\begin{center}
\begin{tabular}{cc}
{\bf(a)}&{\bf(b)}\\
\epsfig{file=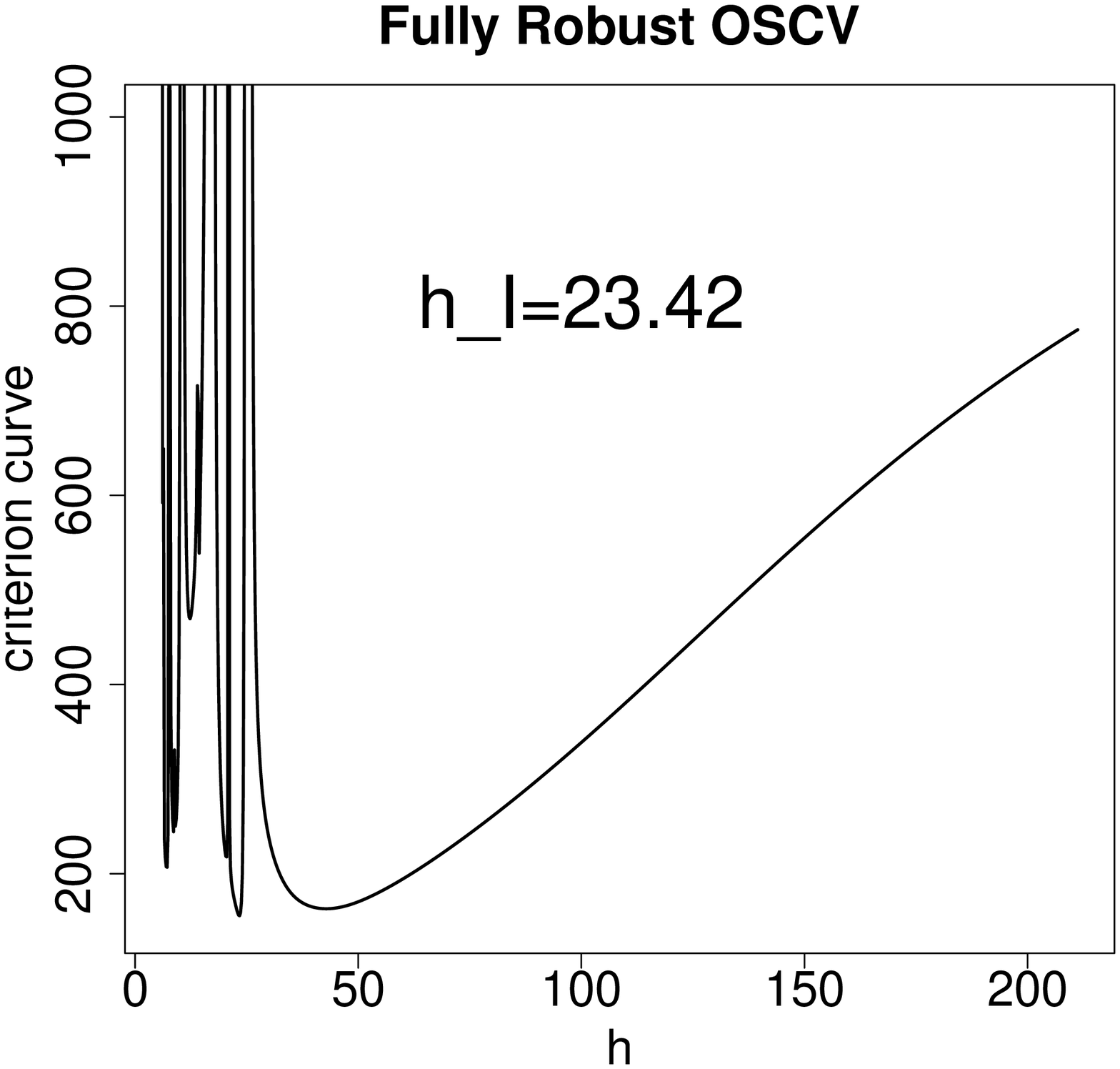,height=200pt}&\epsfig{file=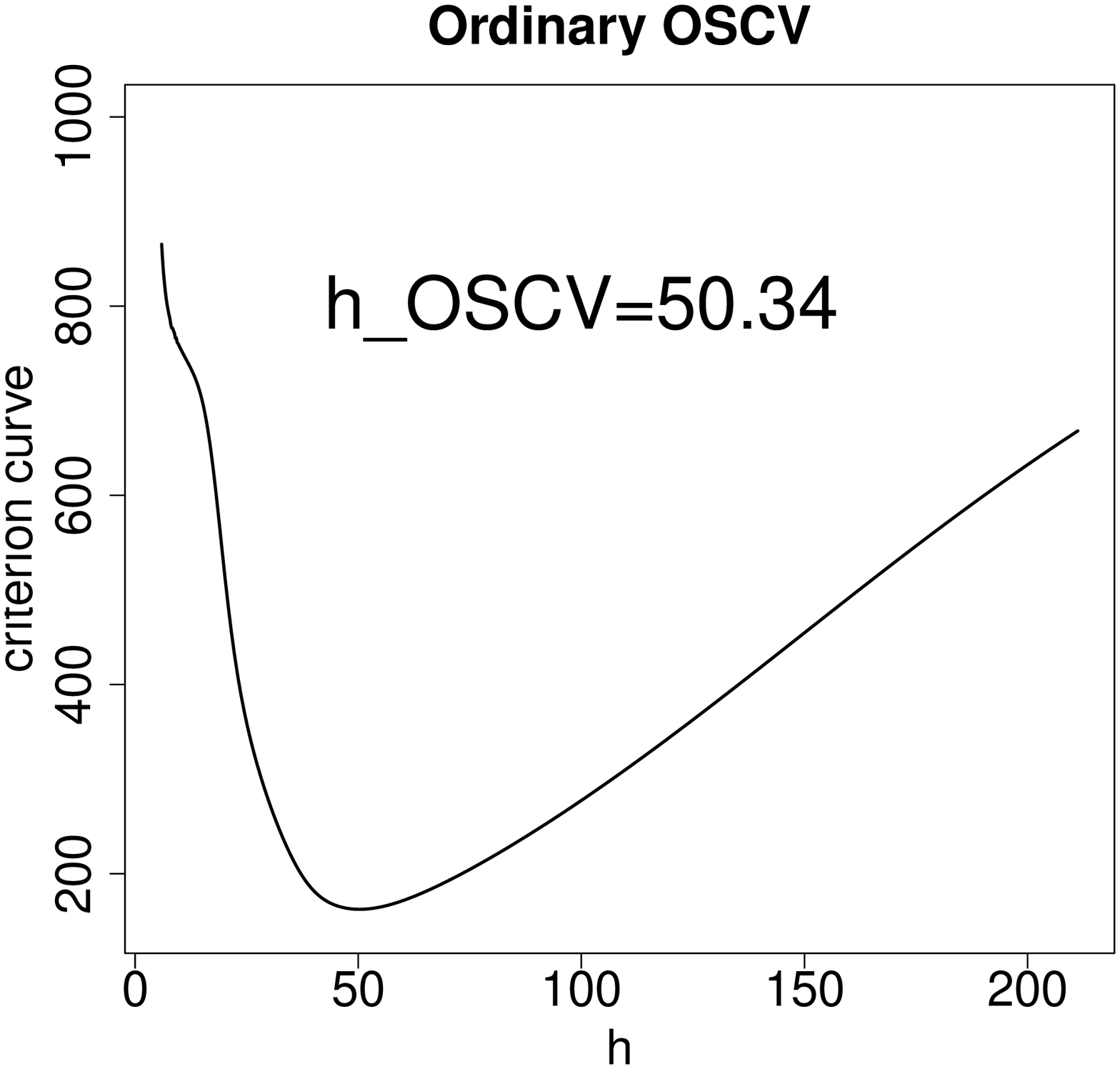,height=200pt}\\
{\bf(c)}&{\bf(d)}\\
\epsfig{file=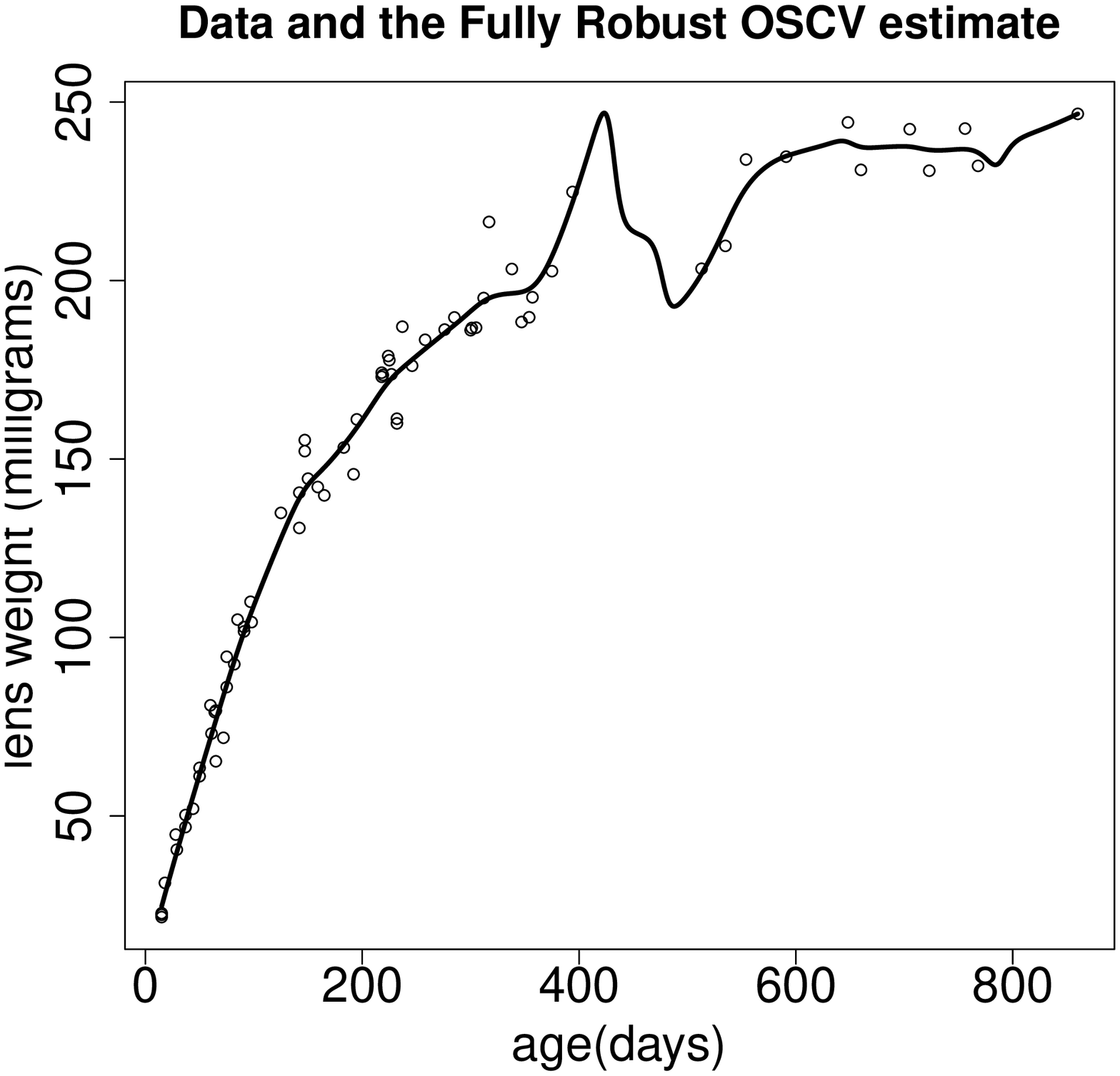,height=200pt}&\epsfig{file=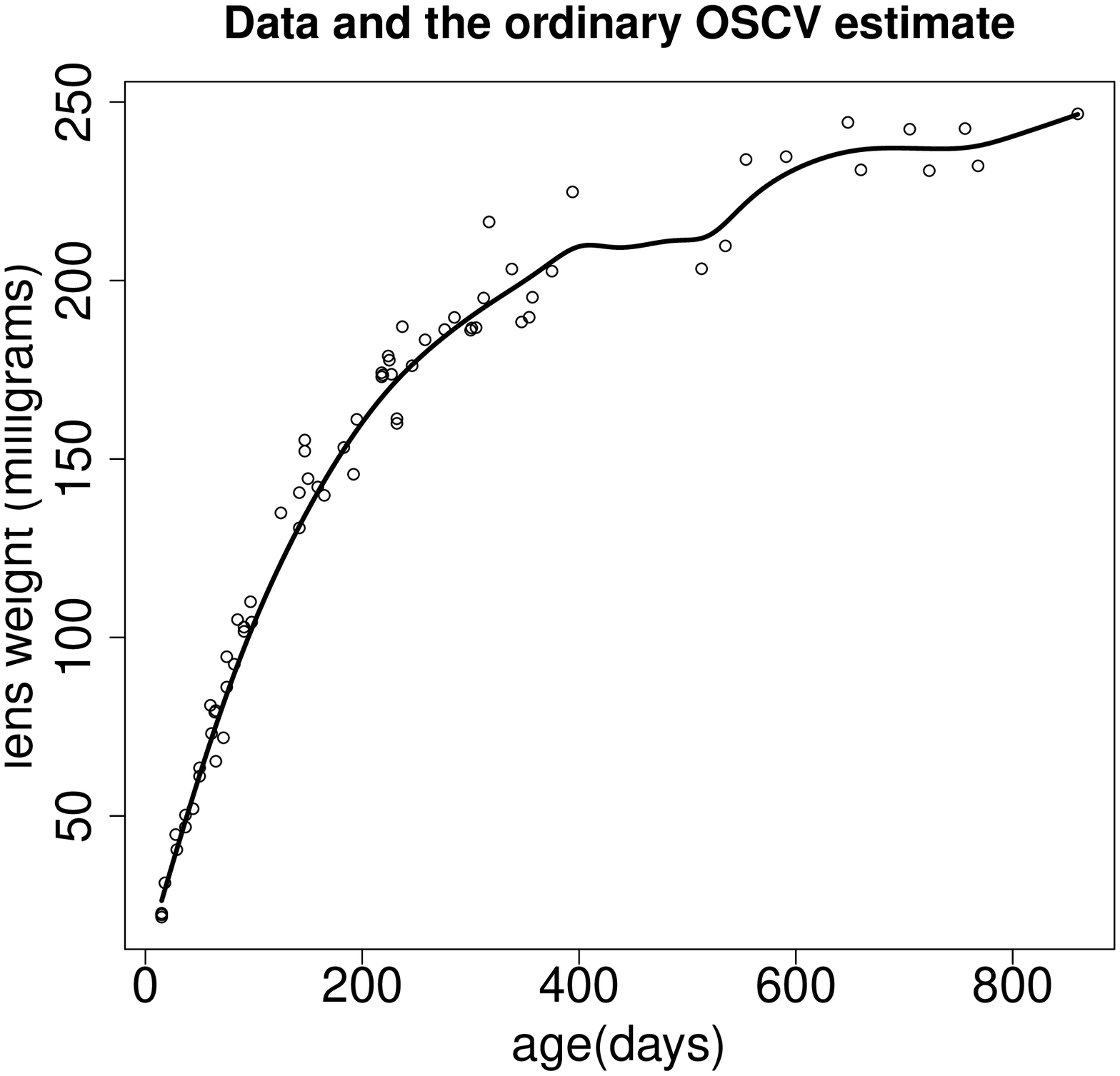,height=200pt}\\
\end{tabular}
\vspace{-0.25cm}
\caption{FROSCV and ordinary OSCV curves and regression fits for the data on eye lens weight and age.\label{fig:rabbit}}
\end{center}
\end{figure}

\noindent \citet{Rabbit} studied the relationship between the eye lens weight and age of rabbits in Australia. The data set of size $n=71$ can be downloaded from~\url"http://www.statsci.org/data/oz/rabbit.html". ~\citet{Rabbit} constructed a model that relates the lens weight ($y$) to age ($x$) as
\[
y=\alpha\mbox{exp}\{-\beta/(x+\gamma)\}
\]
for certain values of $\alpha$, $\beta$, and $\gamma$.
To the contrary of a parametric approach of~\citet{Rabbit}, we estimated $r$ by using the LLE. The table below shows the bandwidths produced for the rabbits' data by different methods.
\begin{center}
\begin{tabular}{|c|c|c|c|}
\hline
$\hat h_{OSCV}$&$\hat h_I$&$\hat h_{CV}$&$\hat h_{PI}$\\
\hline 50.34&23.42&46.95&54.48\\
\hline
\end{tabular}
\end{center}
All methods but fully robust OSCV produce comparable bandwidths and similar regression fits.
Figure~\ref{fig:rabbit} {\bf(a)} and {\bf(b)} shows the $H_I$- and $\phi$-based OSCV curves, correspondingly. For each graph the scale along the horizontal axis is changed such that the global minimum is attained at $\hat h_I=23.42$ in the case of $H_I$ and $\hat h_{OSCV}=50.34$ in the case of $\phi$.
The corresponding local linear estimates are shown in Figure~\ref{fig:rabbit} {\bf(c)} and {\bf(d)}. The fit by ordinary OSCV is quite similar to that obtained by~\citet{Rabbit}. The regression estimate produced by $H_I$ is undersmoothed because of inappropriately small value of $\hat h_I$ obtained from a spurious wiggle of the $H_I$-based OSCV curve. Notice that the largest local minimum of the curve is attained at $h=42.74$ that produces a regression estimate similar to the one corresponding to the $\phi$-based OSCV version. This example and our numerous empirical experience suggest modifying the bandwidth section rule for the $H_I$-based OSCV version so that $\hat h_I$ corresponds to the largest local minimum of the $H_I$-based OSCV curve. This suggestion is similar to that given by~\citet{HallMarron:LocMin} in the context of the kernel density estimation.

\section{Summary and Conclusions}

 The OSCV method is a two-stage procedure. In the first stage one determines the minimizer of the OSCV curve computed based on the kernel $H$ that is, generally, different from the kernel $K$ used in computing the resulting regression estimate $\hat r_h$. The second stage consists in rescaling the bandwidth obtained in the first stage by using the multiplicative constant that is completely determined by $K$, $H$, and the smoothness of a regression function $r$. Unless the smoothness of $r$ is specified, one by default uses the smooth rescaling constant, as this is the case in the original OSCV version of~\citet{HartYi}.

Out of the most often used kernels $K$, such as the Epanechnikov, quartic or Gaussian kernel, the latter one has the largest discrepancy between the smooth and nonsmooth rescaling constants. Thus, for $K=H=\phi$, using the smooth rescaling constant in the case of a nonsmooth function $r$ results in the asymptotic relative bandwidth bias of 16.74\%. Asymptotically, this bias further produces 4.02\% MASE increase. This inspired~\citet{Savchuk:OSCVnonsmooth} to develop the method's correction, termed fully robust OSCV. The idea behind the fully robust OSCV method is to set $K=\phi$ and choose $H$ that produces equal smooth and nonsmooth rescaling constants. Such a kernel $H$ is called robust since it produces consistent OSCV bandwidths regardless of smoothness of $r$.

The current implementation of the fully robust OSCV method is based on the kernel $H_I$ that is drastically close to the Gaussian kernel $\phi$ in a wide range of values of an argument, but has negative tails. Despite this fact, the second moments of $L_\phi$ and $L_I$, the one-sided counterparts of $\phi$ and $H_I$, respectively, are quite different. The discrepancy is caused by different tail behaviours of $L_{\phi}$ and $L_I$. This difference is the main factor that leads to equality of the smooth and nonsmooth rescaling constants in the case of $H_I$.

The practical performances of the $H_I$- and $\phi$-based OSCV versions are compared based on the real data examples and the results of the numerical study of~\citet{Savchuk:OSCVnonsmooth}. For a given data set, the $H_I$- and $\phi$-based OSCV curves are usually quite close, except in the neighborhood of zero, where the $H_I$-based curve might exhibit spurious bumps, that are the artifacts of negativity of the tails of $H_I$. Except for small sample sizes ($n<100$), where the wiggles in the $H_I$-based curve may result in a fake global minimum, the minimizers of the $H_I$-based and $\phi$-based OSCV curves, $\hat b_I$ and $\hat b_{OSCV}$, respectively, are usually about the same in both smooth and nonsmooth cases. 
To avoid the problem of selecting $\hat h_I$ from the ``wiggly part'' of the $H_I$-based OSCV curve that might happen at ``small'' $n$, we suggest that $\hat h_I$ corresponds to the largest $H_I$-based OSCV curve minimizer.

In finite samples, the distribution of the $H_I$-based OSCV bandwidths is usually shifted downwards compared to that of the ASE-optimal bandwidths. However, the magnitude of the relative bandwidth bias by $H_I$ decreases as the smoothness of $r$ decreases. We assessed the absolute value of the relative bandwidth bias produced by $H_I$ and $\phi$ for $100\leq n\leq 1000$. For $H_I$, the absolute value of the relative bandwidth bias is about 13\% in the case of $r_1$, 9\% in the case of $r_2$, and, finally, about 5\% in the case of $r_3$. For $\phi$, the relative bandwidth bias is under 2.1\% in the case of $r_1$, but it exceeds 6\% in the case of $r_2$ and 10\% in the case of $r_3$. In the nonsmooth case, the asymptotically predicted relative bandwidth bias of 16.74\% is not attained by $\phi$ in the considered range of $n$ values, even in the case of the least smooth regression function $r_3$.

The relative bandwidth bias computation and the fact $\hat b_I\approx b_{OSCV}$ suggest that in the smooth case using $H_I$ instead of $\phi$ is practically equivalent to adding wiggles to the OSCV curve along with using a wrong rescaling constant that produces too low bandwidth. There is some benefit of using $H_I$ in the case where $r$ has multiple cusps, though. However, since the nonsmooth Gaussian constant $C_{\phi}^*$ is about equal to $C_I$, we suggest that in the case when nonsmoothness of $r$ is evident from the scatter diagram of the data, one uses $\phi$ along with $C_{\phi}^*$ instead of $H_I$ along with $C_I$. The benefit of the former combination over the latter one is obtaining a smoother OSCV curve.

The kernel $H_I$ uncovers the OSCV method's sensitivity to insignificant modifications of the kernel used in the cross-validation stage. Indeed, tiny deviation of $H_I$ from $\phi$ in the tails greatly changes theoretical properties and practical performance of the OSCV method.

Nonnegative robust kernels are expected to produce smoother OSCV curves compared to the negative-tailed kernel $H_I$. Our search for nonnegative robust kernels resulted in the bimodal kernel $H_B$ and several other robust bimodal kernels. Even though $H_B$ yields smoother OSCV curves compared to $H_I$, we found that bimodality of $H_B$ is associated with producing the curves with multiple local minima. This encourages a new search for {\em nonnegative unimodal} robust kernels in the case $K=\phi$.


\bibliographystyle{abbrvnat}
\bibliography{../../../refs}
\section*{Appendix}

Regression functions $r_1$, $r_2$, and $r_3$ are defined below. For each function, $0\leq x\leq 1$.

\[
\begin{array}{l}
\displaystyle{r_1(x)=5x^{10}(1-x)^2+2.5x^2(1-x)^{10},}\\\\
\displaystyle{r_2(x)=\begin{cases}
0.0125-0.05|x-0.25|,\qquad\qquad 0\leq x\leq0.5,\\[0.15cm]
0.05|x-0.75|-0.0125,\qquad 0.5<x\leq 1.
\end{cases}}\\\\
\displaystyle{
r_3(x)=\begin{cases}
0.047619\sqrt x,\qquad 0\leq x<0.1,\\
0.035186e^{-20x}+0.010297,\qquad 0.1\leq x<0.3,\\
0.142857x-0.032473,\qquad 0.3\leq x<0.35,\\
0.142857(x-0.35)(x-0.45)+0.017527,\qquad 0.35\leq x<0.6,\\
0.151455-0.214286x,\qquad 0.6\leq x<0.7,\\
0.001455-0.214286(x-0.7)^3(x-0.4),\qquad 0.7\leq x<0.8,\\
0.004762\ln(10x-7.9)+0.012334,\qquad 0.8\leq x\leq 1.
\end{cases}
}
\end{array}
\]

\end{document}